\documentclass[sigconf, nonacm]{acmart}

\usepackage{amsmath}
\usepackage{algorithm}
\usepackage{algpseudocode}
\usepackage{float}
\usepackage{subcaption}
\usepackage{multirow}
\usepackage{hhline}
\usepackage{makecell}
\usepackage{booktabs}
\usepackage{pifont}
\usepackage{enumitem}

\algrenewcommand\algorithmicrequire{\textbf{Input:}}
\algrenewcommand\algorithmicensure{\textbf{Output:}}

\begin{document}

\title{Rimuru: Unified Memory Management for Modern LLM Serving via Elastic Expert Offloading}

\author{Zhiyuan Fang}
\affiliation{
  \institution{Sun Yat-sen University}
  \city{Zhuhai}
  \country{China}
}
\email{fangzhy27@mail2.sysu.edu.cn}

\author{Dingyang Li}
\affiliation{
  \institution{Sun Yat-sen University}
  \city{Zhuhai}
  \country{China}
}
\email{lidy63@mail2.sysu.edu.cn}

\author{Jianfeng Zhong}
\affiliation{
  \institution{Sun Yat-sen University}
  \city{Zhuhai}
  \country{China}
}
\email{zhongjf25@mail2.sysu.edu.cn}

\author{Bin Wang}
\affiliation{
 \institution{Independent Researcher}
 \city{Shanghai}
 \country{China}
}
\email{bw373717@gmail.com}

\author{Weixi Zhu}
\affiliation{
 \institution{Independent Researcher}
 \city{Shanghai}
 \country{China}
}
\email{weixi.zhu@openeuler.sh}

\author{Cunshu Ni}
\affiliation{
 \institution{Independent Researcher}
 \city{Hangzhou}
 \country{China}
}
\email{cunshuni200@gmail.com}

\author{Zicong Hong}
\affiliation{
  \institution{EPFL}
  \city{Lausanne}
  \country{Switzerland}
}
\email{zicong.hong@epfl.ch}

\author{Wuhui Chen}
\affiliation{
  \institution{Sun Yat-sen University}
  \city{Zhuhai}
  \country{China}
}
\affiliation{
  \institution{Peng Cheng Laboratory}
  \city{Shenzhen}
  \country{China}
}
\email{chenwuh@mail.sysu.edu.cn}

\author{Zibin Zheng}
\affiliation{
  \institution{Sun Yat-sen University}
  \state{Zhuhai}
  \country{China}
}
\email{zhzibin@mail.sysu.edu.cn}

\renewcommand{\shortauthors}{Zhiyuan Fang et al.}

\begin{abstract}
Modern LLMs strain device memory through large model weights and attention states, particularly in resource-constrained scenarios. Existing systems separately provision memory for model weights and attention states, imposing fixed fixed boundaries that cannot adapt to the changing memory demands of workloads. Under heavy workloads, insufficient state capacity limits request concurrency and reduces throughput; under light workloads, idle memory in one region cannot be reallocated to another, wasting device memory. Meanwhile, modern LLM architectures introduce heterogeneous tensors with diverse layouts and lifetimes, further complicating memory management. These problems motivate unified memory management for heterogeneous tensors. However, a shared pool alone does not make memory directly reusable across tensors with different allocation granularities and execution layouts.

This paper presents Rimuru, a modern LLM serving system that unifies memory management through elastic expert offloading. 
Rimuru establishes a shared page pool to enables elastic memory sharing across heterogeneous tensors.
To accommodate different allocation granularities, Rimuru introduces memory views that allow each tensor type to interpret the shared memory space using its own page size and page table. Furthermore, Rimuru combines huge-page alignment with a multi-level page allocator that splits and merges pages on demand, enabling cross-type capacity reuse while minimizing fragmentation. A page-table-aware fused MoE kernel directly accesses non-contiguous expert weights, avoiding copies into contiguous execution buffers. Building on these mechanisms, Rimuru employs a workload-aware adjuster that dynamically reallocates pages across tensor types according to runtime demands, while coordinating changes in expert residency with the prefetching and execution pipeline. Implemented on top of vLLM, Rimuru remains compatible with modern LLMs. Evaluations show that Rimuru achieves up to 19.03$\times$ baseline throughput and reduces TTFT by 98.1\% on average.

\end{abstract}


\maketitle

\section{Introduction}

Large Language Models (LLMs), owing to their remarkable capabilities, have been increasingly penetrated into workflows across a wide range of domains~\cite{team2025kimi, liu2025deepseek, minimax25}. Recently, a growing trend toward on-premises and private deployment has emerged. Motivated by the demand for data privacy, customized experiences, and low latency, many small and medium-sized enterprises (SMEs) are increasingly adopting locally deployed LLMs to support internal applications~\cite{pan2025measuring}, such as code agents, enterprise chatbots, and document processing. Meanwhile, individual developers have also shown rising interest in building personal AI assistants for everyday productivity and development tasks~\cite{OpenClaw, HermesAgent}.

\begin{figure}[t]
  \centering
  \includegraphics[width=\linewidth]{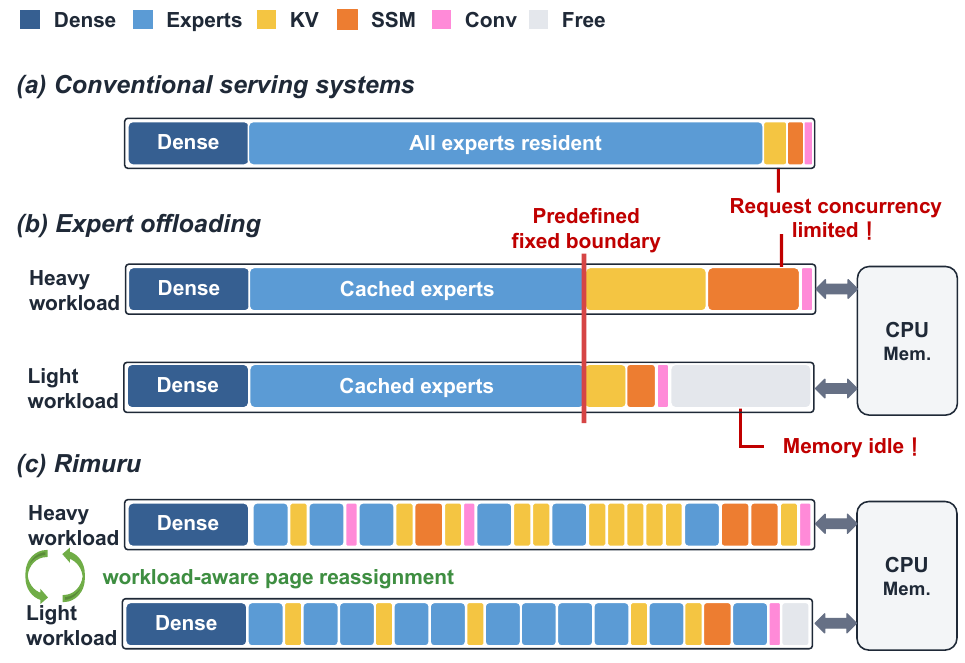}
  \caption{Different memory management. (a) Conventional serving systems leave little memory for attention states. (b) Expert offloading creates headroom but still exist a fixed boundary between the expert cache and attention states. (c) Rimuru with unified memory management.}
  \label{figure:memory_management_concept}
\end{figure}

However, due to cost constraints, limited device memory remains a critical bottleneck~\cite{song2024powerinfer, sheng2023flexgen, alizadeh-etal-2024-llm}. Conventional serving systems keep the complete model weights on the devices, which can exhaust the available memory. 
What's worse, this affects more than whether a model can be loaded. 
Serving requests retain intermediate states across forward steps, and batching multiple requests requires substantial device memory, especially for modern applications with long context lengths of up to 1M tokens~\cite{10.1145/3694715.3695948}.
As shown in~\autoref{figure:memory_management_concept}(a), the remaining memory space for attention states is limited, making it difficult to support large-batch inference and resulting in low throughput. 
This issue is particularly pronounced as most modern large-scale models adopt the MoE architecture, which provides strong performance but incurs substantial memory demands. 
For example, deploying DeepSeek-V4-Flash~\cite{xu2026deepseek} on eight NVIDIA A100 GPUs leaves only about 7 GB of device memory per GPU for attention states, sufficient to support just two concurrent requests with 1M-token contexts.

To address memory constraints, expert offloading provides an effective solution~\cite{xue2024moe, fang2026fate, yu2025fmoe, chen2025ktransformers}. 
These approaches leverage the sparse activation property of experts, retain a subset of experts on the device and keep the remaining experts in host memory.
Then through expert caching, prediction, or CPU--GPU co-execution, they reduce or hide on-demand loading overhead. 
As a result, as shown in~\autoref{figure:memory_management_concept}(b), the reduced weight footprint creates additional capacity for larger batch sizes and can improve serving throughput.
However, existing approaches still lack sufficient elasticity. 
Modern LLM serving systems commonly adopt PagedAttention, which manages KV caches through paged allocation within a dedicated memory region while keeping model weights outside the managed memory pool~\cite{kwon2023pagedattention, zheng2024sglang, zhang2025jenga}. 
Such memory isolation fails to accommodate the dynamic memory demands of serving workloads~\cite{sun2024llumnix}, resulting in low memory utilization.
As illustrated in~\autoref{figure:memory_management_concept}(b), this design maintains a predefined fixed boundary between different tensor types. 
During workload bursts, attention states cannot reclaim capacity held by cached experts that could otherwise be evicted, restricting request admission. 
Under light workload, idle state-pool capacity cannot retain additional experts to reduce weight transfers. 
We further investigate the impact of this limitation on serving performance in Section~\ref{subsec:fundamental_mismatch}, where our case study shows that elastic memory sharing further reduces total workload completion step by approximately 30\%.
This situation thus reveals a broader problem: heterogeneous tensors should share device memory elastically under dynamic workloads.

More broadly, this problem intensifies as modern LLM architectures become increasingly heterogeneous. Recent models combine MoE layers~\cite{fedus2022switch, jiang2024mixtral}, shared and fine-grained experts~\cite{dai2024deepseekmoe, yang2025qwen3, muennighoff2024olmoe}, and linear-attention or Mamba layers~\cite{gu2023mamba, lieber2024jamba, qwen35blog}. Consequently, memory management for modern LLMs must accommodate tensors with increasingly diverse layouts, lifetimes, and access patterns, rather than one dominant state such as the KV cache.

A natural way to break this isolation and improve memory utilization is to manage heterogeneous tensors in a unified memory space, dynamically adjusting capacity for each tensor type. Memory released by one tensor type must be directly reusable by another without costly tensor reorganization. 
However, enabling such sharing introduces three key challenges.
First, heterogeneous tensors require different allocation granularities and indexing schemes. Small pages support fine-grained attention-state allocation but increase mapping overhead for large experts, whereas large pages waste capacity on small states. 
Second, page-based storage can distribute cached experts across separate pages, whereas conventional fused MoE kernels expect a regular, contiguous expert-weight layout. Gathering their weights into contiguous execution buffers restores the expected layout but adds local copies overhead. The execution path must therefore locate each routed expert across these pages without repeatedly repacking weights.
Third, adjusting the expert cache requires not only adapting to changing workload demands but also balancing request concurrency against weight-loading overhead. Increasing concurrency requires reclaiming memory from cached experts, but excessive expert eviction incurs substantial I/O overhead. Moreover, page reclamation and expert residency updates must be coordinated with asynchronous prefetching and execution to avoid cache oscillation, redundant data transfers, and I/O contention.

To address these challenges, we present Rimuru, a modern LLM serving system with unified memory management designed to improve throughput in resource-constrained environments. The key idea of Rimuru is to manage device memory in a shared pool across heterogeneous tensors and dynamically adjust memory allocation through elastic expert offloading, maximizing the utilization of expensive device memory.

Specifically, Rimuru introduces memory views to decouple shared physical memory from tensor-specific allocation and addressing. 
Each tensor type interprets the underlying storage through its own memory view, with a dedicated page size and page table. 
This multi-view memory layout accommodates different tensor shapes and sizes without imposing a uniform allocation granularity. 
Rimuru further defines a huge-page size as the least common multiple of the page sizes across expert and SSM, providing aligned boundaries that prevent overlap between pages of different tensor types. 
To enable memory sharing, Rimuru employs a multi-level page allocator that organizes pages by size and supports on-demand allocation and reclamation. This design avoids assigning unnecessarily large pages to fine-grained tensors, thereby reducing internal fragmentation.
To preserve execution efficiency, Rimuru extends the fused MoE kernel with page-table indirection, allowing it to access expert weights directly from cache pages or loading buffers without copying them into contiguous execution buffers.
Building on unified memory management, Rimuru employs a workload-aware expert cache adjuster that dynamically changes the expert cache size in response to workload demands. During request bursts, it reduces expert residency to free memory for additional concurrent requests, improving throughput. As memory becomes available, it restores cached experts to reduce CPU--GPU data transfers and serving latency.
As illustrated in~\autoref{figure:memory_management_concept}(c), Rimuru elastically reallocates memory across heterogeneous tensors to adapt to changing workloads, improving memory utilization and serving performance.

To summarize, we make the following contributions:
\begin{itemize}
  \item We identify growing tensor heterogeneity, and a fundamental mismatch between dynamic memory demand and static memory isolation in modern LLM serving systems.
  \item We design and implement Rimuru, which combines tensor-specific memory views, multi-level page allocation, and page-table-aware expert execution to support efficient cross-type memory reuse. Building on this unified memory substrate, Rimuru elastically adjusts expert residency according to runtime memory demands while coordinating asynchronous expert loading and execution.
  \item We conduct comprehensive experiments across diverse platform scale, models and workloads. Compared with baselines, Rimuru achieves up to 19.03$\times$ throughput and reduces TTFT by 98.1\% on average.
\end{itemize}





\section{Background and Motivation}

\subsection{Modern LLM Architectures and Tensor Heterogeneity}


\begin{figure}[t]
    \centering
    \subfloat[Classic.\label{fig:transformer}]{%
        \includegraphics[width=0.3\linewidth]{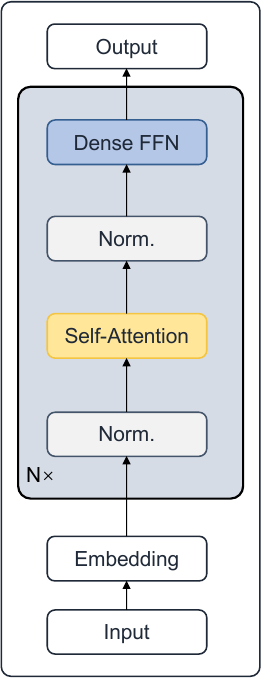}
    }\hfill
    \subfloat[Modern LLM.\label{fig:modernllm}]{%
        \includegraphics[width=0.672\linewidth]{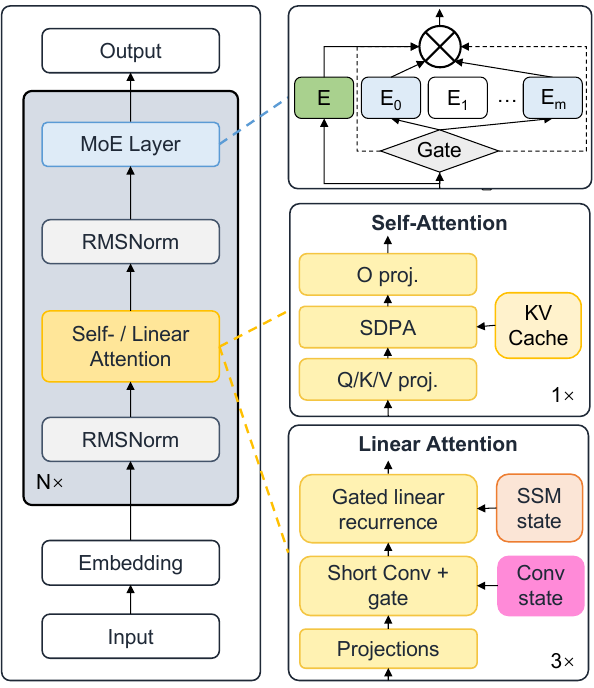}
    }
    \caption{A comparison of the architecture of a classic transformer-based LLM and a representative modern LLM.}
    \label{figure:modern_architecture}
\end{figure}

The conventional dense Transformer architecture consists of repeated blocks of self-attention and dense FFN modules, as illustrated in~\autoref{figure:modern_architecture}(a). 
Its primary dynamically managed state during inference is the KV cache.
To scale model capacity without proportionally increasing computation, modern LLMs increasingly adopt MoE layers~\cite{fedus2022switch, jiang2024mixtral}. Recent architectures further incorporate shared experts and fine-grained expert partitioning to improve expert specialization~\cite{dai2024deepseekmoe, yang2025qwen3, muennighoff2024olmoe}.
Meanwhile, the demand for long-context inference has driven the adoption of linear attention and Mamba layers, which introduce persistent runtime states beyond the KV cache, including State Space Model (SSM) and convolution states~\cite{gu2023mamba, lieber2024jamba, qwen35blog}.
As a result, modern LLMs increasingly combine heterogeneous layers rather than relying on a uniform Transformer architecture, as illustrated in~\autoref{figure:modern_architecture}(b).
This architectural evolution introduces tensors with diverse memory requirements, making efficient memory management increasingly challenging, particularly in resource-constrained environments.

\subsection{Memory Management in LLM Serving Systems}

\textbf{Attention-State-centric Memory Management.} We use \textit{attention states} to refer to all persistent, per-request states maintained during inference, including the KV cache, as well as the SSM and convolution states in Mamba or hybrid attention models~\cite{lieber2024jamba,qwen35blog}.
Mainstream serving systems target large-scale distributed deployment, where attention states dominate device memory under high concurrency. Their memory management thus centers on attention states.
To improve efficiency, modern frameworks adopt PagedAttention, which organizes attention states into pages allocated on demand from a global pool~\cite{kwon2023pagedattention, zhang2025jenga}, thereby largely eliminating fragmentation.
Building on this, RadixAttention~\cite{zheng2024sglang} exploits prefix sharing across requests by organizing KV cache in a radix tree, enabling efficient reuse and reducing redundant computation. 
PD-disaggregation frameworks~\cite{patel2024splitwise, zhong2024distserve, qin2026mooncake, xie2026strata} further externalize attention states storage into a disaggregated memory pool accessible across prefill and decode stages, allowing KV cache to be managed, placed, and reused independently across stages and devices.
Clearly, existing serving systems place strong emphasis on paged management of attention states. At the same time, they seldom explicitly manage the memory of model weights, which leads to suboptimal performance in resource-constrained environments, where model weights dominate GPU memory.

\textbf{Expert-centric Memory Management.} 
Expert offloading systems~\cite{fang2026fate,yu2025fmoe,xue2024moe} are designed to serve MoE models in resource-constrained environments, where the large memory footprint of expert weights poses a major bottleneck. These systems therefore focus on expert memory management, primarily employing expert caching and prediction to reduce GPU-CPU data transfer overhead. Because expert activation patterns are often skewed, frequently activated experts are preferentially retained in GPU memory, with many systems adopting LRU-based policies for cache replacement.
In contrast to caching, expert prediction is more challenging because expert activation depends on gating decisions that are difficult to predict~\cite{hwang2024pregatedmoe, tang2026moe}. Nevertheless, prior studies have made substantial progress in improving prediction accuracy and reducing offloading overhead.
Despite these advances, existing offloading systems primarily optimize expert memory management without jointly managing expert weights and attention states. This separation prevents memory capacity from being dynamically shared across tensor types, leading to inefficient device memory utilization.
Unlike these systems, Rimuru enables unified memory management across heterogeneous tensors, allowing device memory to be dynamically reallocated according to workload demands. This is the key difference between Rimuru and prior work.

\subsection{A Fundamental Mismatch: Dynamic Memory Demand vs. Static Memory Isolation}
\label{subsec:fundamental_mismatch}



\begin{figure}[t]
    \centering
    \subfloat[Dynamism of request arrival.\label{fig:Dynamics_a}]{%
        \includegraphics[width=0.48\linewidth]{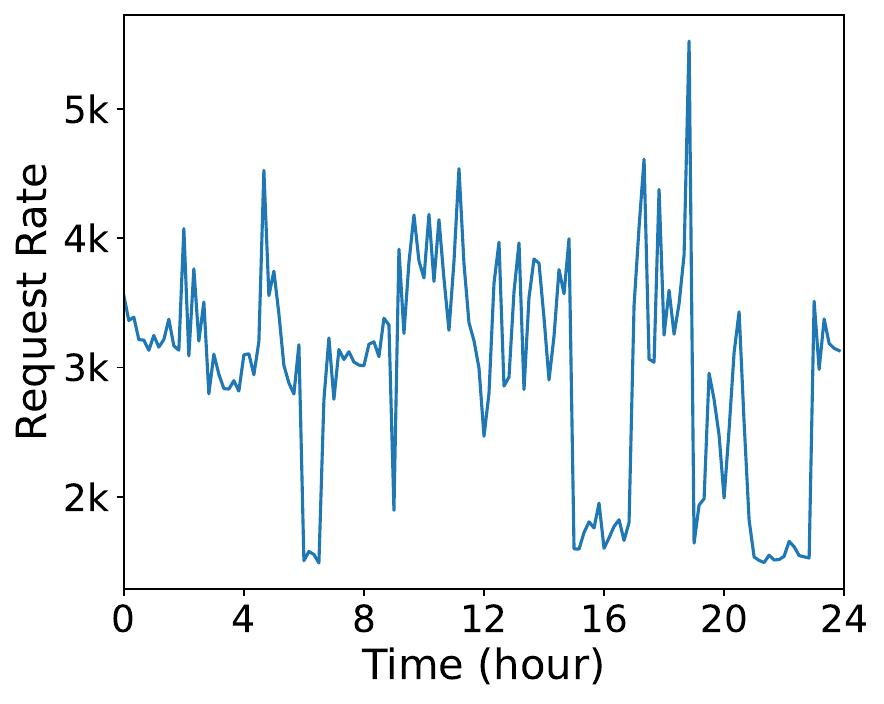}
    }\hfill
    \subfloat[Dynamism of request length. p$N$ means the lengths of $N\%$ requests $\leqslant$ the corresponding y-axis value.\label{fig:Dynamics_b}]{%
        \includegraphics[width=0.48\linewidth]{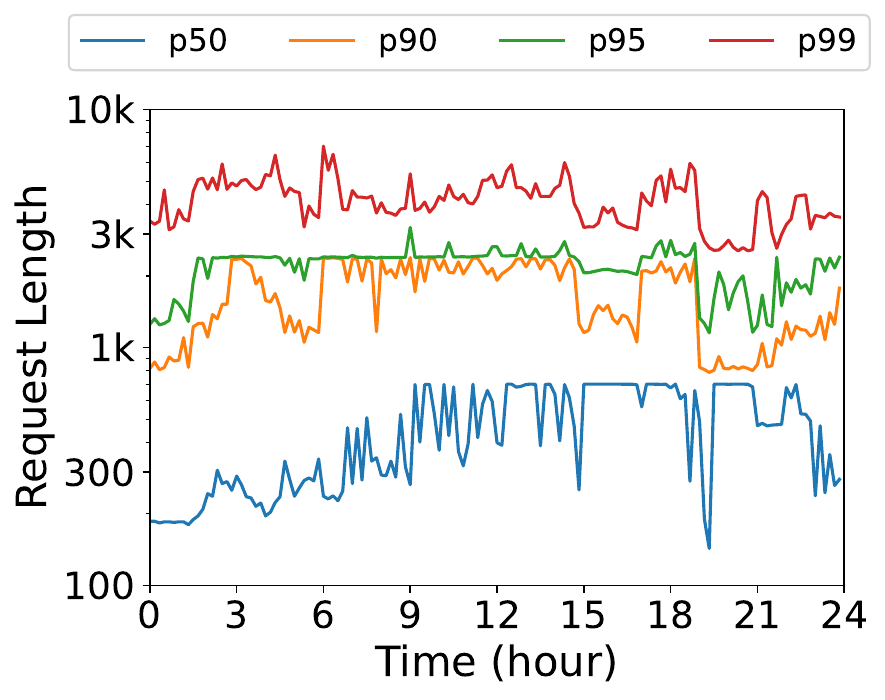}
    }

    \caption{Request arrival rate and sequence length vary over time in a production workload.}
    \label{figure:Dynamics}
\end{figure}

During MoE inference, memory demand is dynamic~\cite{sun2024llumnix, 10.1145/3711896.3737413}.
First, workloads change over time, and both request arrival patterns and sequence lengths can vary widely in practice~\cite{yu2022orca, zhong2024distserve}.
\autoref{figure:Dynamics} illustrates a representative example from a real production environment, where bursty traffic and long requests can abruptly increase memory demand~\cite{xiang2026servegen}. 
Second, the memory requirements of different tensor types vary across inference phases. During the prefill stage, the model processes many tokens and activates nearly all experts, dominating memory usage, while the KV cache grows rapidly. During decoding, the number of active experts decreases, while the KV cache persists and continues to grow.
These dynamics indicate that optimal device memory allocation must adapt continuously over time.

\begin{figure}[t]
    \centering
    \subfloat[vLLM: resident model weights leave limited capacity for the KV cache. The maximum request concurrency is limited.\label{fig:case_study_a}]{%
        \includegraphics[width=\linewidth]{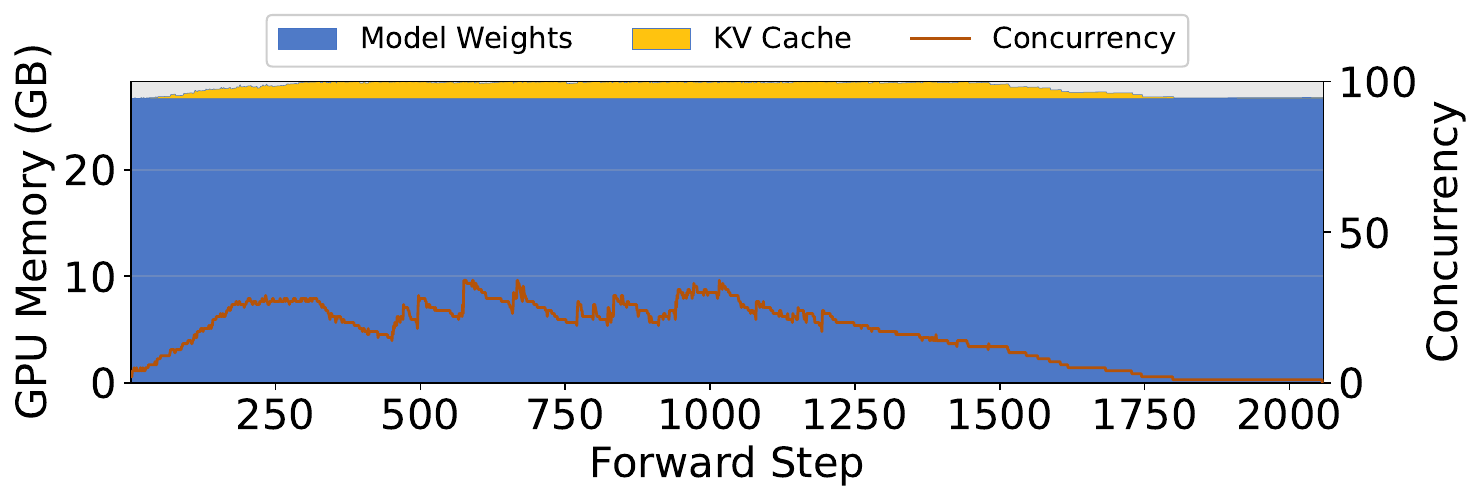}
    }
    
    \subfloat[Expert offloading on vLLM: a fixed expert cache leaves more, but still bounded, capacity for the KV cache.
    \label{fig:case_study_b}]{%
        \includegraphics[width=\linewidth]{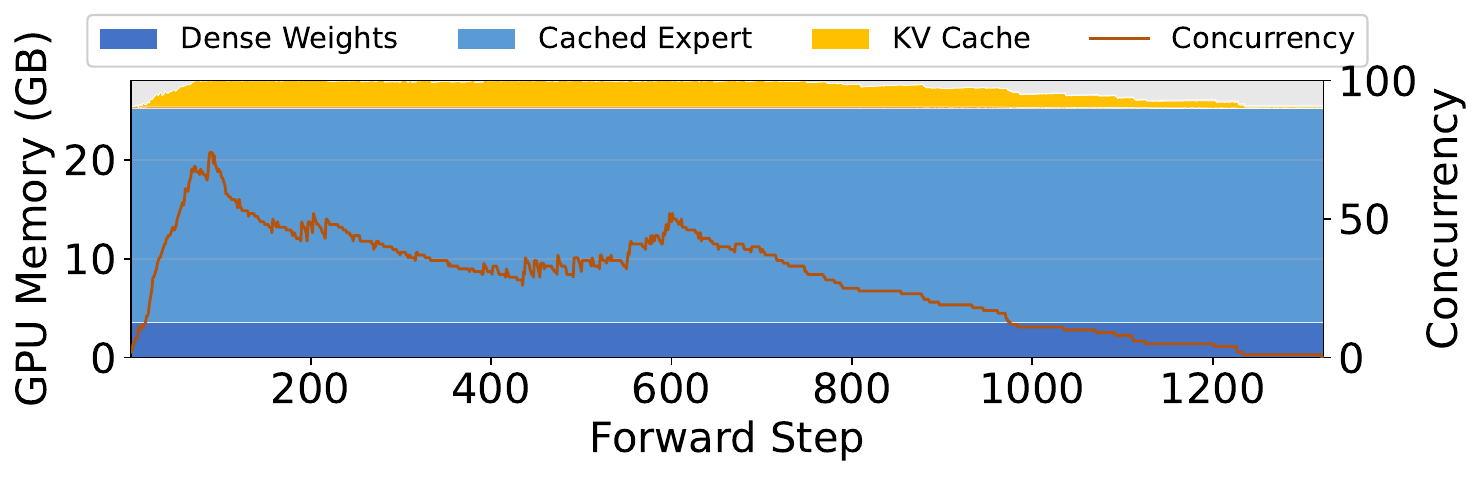}
    }

    \subfloat[Rimuru: pages are allocated dynamically between the KV cache and experts.\label{fig:case_study_c}]{%
        \includegraphics[width=\linewidth]{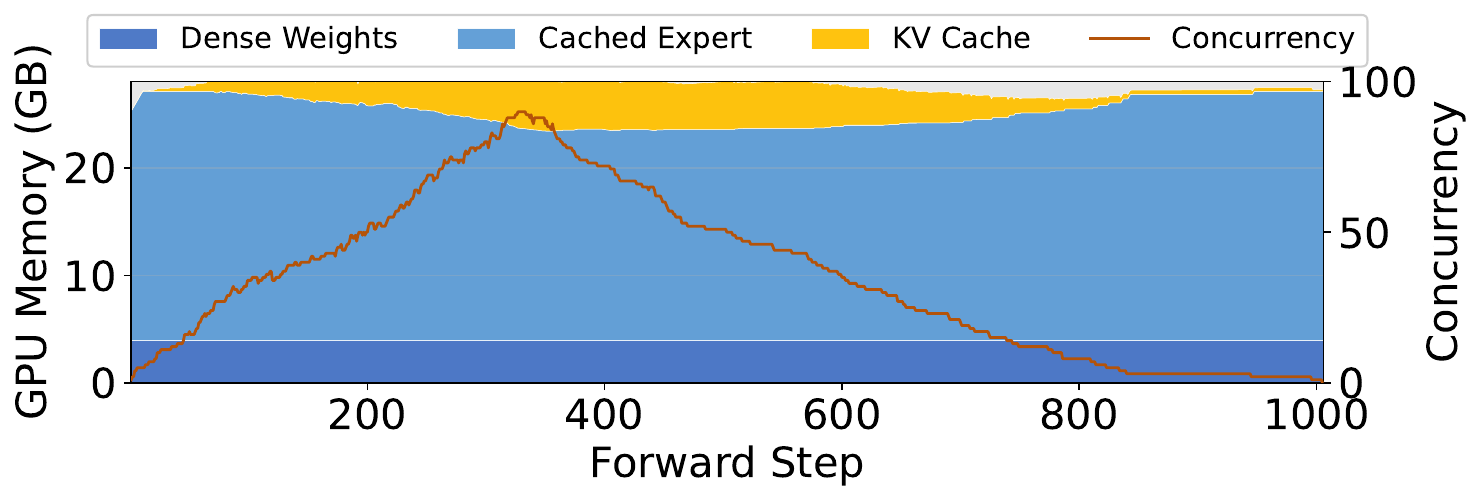}
    }

    \caption{GPU-memory usage and request concurrency while serving Qwen1.5-MoE-14B on one NVIDIA RTX 5090. All configurations serve the same ShareGPT requests~\cite{ShareGPT}, submitted at 10 requests/s, with GPU memory utilization capped at 0.9. Stacked areas show memory attributed to model weights and the KV cache; the line shows running-request concurrency. The horizontal axis reports forward steps.}
    \label{figure:case_study}
\end{figure}

However, existing serving systems still manage expert caches and attention states in isolated memory regions, which fails to accommodate their dynamic memory demands. 
First, modern serving systems commonly adopt PagedAttention, which reserves a dedicated memory region for attention states and manages it through fine-grained page allocation. Although this design reduces fragmentation, it prevents memory sharing across tensor types, leading to memory underutilization. 
Second, existing expert offloading strategies typically employ fixed offloading ratios, lacking the flexibility to adapt to changing workloads. 
Under light workloads, unused memory in the attention-state pool cannot be used to cache additional experts and reduce weight-loading overhead. 
Under heavy workloads, the attention-state pool cannot expand by evicting cached experts, forcing requests to wait even when some expert-cache could be evicted. 
This inefficiency is particularly pronounced in resource-constrained environments, where device memory is scarce. 
These limitations motivate breaking memory isolation and enabling elastic memory allocation across heterogeneous tensors to improve memory utilization.


\textbf{Case study.} To quantify this mismatch, \autoref{figure:case_study} compares three progressively enhanced designs under the same model, memory budget, and request stream. Case (a) is vLLM~\cite{kwon2023pagedattention}, which is a representative conventional in-GPU serving system with PagedAttention. Case (b) represents an expert offloading system that implements the state-of-the-art offloading strategy of Fate~\cite{fang2026fate} on top of vLLM. Case (c) represents Rimuru, which further enables unified memory management by elastic expert offloading. 

In~\autoref{figure:case_study}(a), resident model weights occupy most of the GPU memory budget, leaving limited capacity for the KV cache. Peak concurrency reaches only 34 requests, and completing the workload requires approximately 2100 forward steps. Although some experts remain inactive during many of these steps, their occupied memory cannot be reclaimed to increase batch size due to the predefined memory boundary.
In~\autoref{figure:case_study}(b), expert offloading frees GPU memory for a larger KV-cache pool, increasing peak concurrency to 74 requests and reducing the required forward steps to approximately 1300. However, the expert-cache capacity remains fixed regardless of workload changes. Under high workload, the KV-cache pool cannot expand by reclaiming memory from cached experts. Under low workload, unused KV-cache capacity cannot be used to retain additional experts and reduce weight-loading overhead. Thus, expert offloading alleviates the memory pressure from model weights but retains the rigid memory boundary between expert caches and attention states, limiting memory utilization.

In~\autoref{figure:case_study}(c), Rimuru adjusts expert residency to accommodate changing KV-cache demand. As concurrency grows, the expert-cache footprint contracts, releasing capacity for the KV cache. As requests complete and KV pages are freed, Rimuru restores cached experts using the returned capacity. The observed peak concurrency reaches 90, and the workload spans about 1000 forward steps. The changing boundary between cached experts and the KV cache in figure shows how shared capacity follows attention-state demand.

This case study reveals the memory waste caused by static memory isolation and its impact on serving performance. It also highlights the distinction between freeing device memory and enabling its dynamic reuse across heterogeneous tensors. Expert offloading provides additional capacity for attention states, whereas elastic sharing allows the division between expert caching and the KV cache to change with demand. In this workload, Rimuru supports higher peak concurrency and processes the request stream in fewer forward steps.

\section{Challenges}

Breaking static memory isolation requires more than placing expert weights and attention states in a shared memory pool. Released capacity must be reusable across tensor types with different allocation and addressing requirements, expert computation must remain efficient as residency changes, and cache resizing must be coordinated with ongoing execution. These requirements introduce three challenges.

\begin{table}[t]
  \centering
  \caption{Tensor heterogeneity in representative modern LLMs. Shapes are reported for each expert‘s three projection matrix, the key and value tensors for one token, and an individual SSM or convolutional state.}
  \resizebox{\linewidth}{!}{
  \footnotesize
  \renewcommand{\arraystretch}{1.2}
  \begin{tabular}{cccc}
    \Xhline{1pt}
    \multicolumn{2}{c}{\multirow{2}{*}{\textbf{Tensor Type}}} & \multicolumn{2}{c}{\textbf{Tensor Shape (single)}} \\
    \cline{3-4}
    \multicolumn{2}{c}{} & \textbf{Qwen3-30B-A3B} & \textbf{Qwen3.5-35B-A3B} \\
    \Xhline{1pt}
    \multirow{2}{*}{\textbf{Expert}} & \textbf{w1, w3} & [2048, 768] & [2048, 512] \\
    \cline{2-4}
    & \textbf{w2} & [768, 2048] & [512, 2048]\\
    \hline
    \multicolumn{2}{c}{\textbf{KV cache}} & [2, 4, 128] & [2, 2, 256] \\
    \hline
    \multirow{2}{*}{\textbf{Mamba state}} & \textbf{conv} & / & [3, 8192]  \\
    \cline{2-4}
    & \textbf{SSM} & / & [32, 128, 128]  \\
    \Xhline{1pt}
  \end{tabular}
  }
  \label{table:heterogeneous_tensor}
\end{table}


\textbf{1) Allocation Granularities and Capacity Reuse.}
Expert weights and attention states differ in both storage structure and allocation granularity, as illustrated by the tensor shapes in~\autoref{table:heterogeneous_tensor}. Expert weights comprise large matrices, KV-cache capacity is allocated incrementally as sequences grow, and SSM and convolutional states require differently shaped per-request storage. A uniform small page size increases mapping and management overhead for large weights, whereas a large page size wastes capacity on small states. Allowing different page sizes addresses this tension but introduces another requirement: capacity released in one granularity must become allocatable in another without overlapping live allocations or leaving free space stranded in incompatible units. The memory manager must therefore preserve tensor-specific allocation and indexing while supporting capacity reuse across tensor types.

\textbf{2) Efficient Expert Execution under Changing Residency.}
Conventional fused MoE kernels typically address routed experts through a regularly packed expert-weight layout. With shared allocation, cached experts can occupy separate page locations, and their mappings change as experts enter or leave the cache. Gathering these weights into a contiguous execution buffer restores the expected layout but adds local device-to-device copies and requires staging capacity, weakening the benefit of caching. The execution path must locate each routed expert in cache pages or loading buffers and consume its weights directly, while preserving the matrix layout required for efficient computation.


\textbf{3) Workload-aware and Execution-safe Cache Resizing.}
Expert-cache capacity determines a trade-off between request concurrency and weight-loading overhead. 
Reclaiming cached experts creates room for attention states, but excessive eviction increases subsequent transfers; retaining too many experts can restrict request admission. 
Resizing decisions must respond to memory pressure without repeatedly evicting and reloading experts as demand fluctuates. 
Applying these decisions also requires coordination with asynchronous prefetching and execution. 
Cache expansion can duplicate or contend with pending prefetches, while newly loaded weights must be ready before their mappings become visible to computation. When shrinking the cache, pages cannot be reassigned while an ongoing forward pass may still access them. 
Cache resizing must therefore coordinate capacity decisions, data transfers, mapping updates, and safe page reclamation.

\section{System Overview}

\begin{figure}
  \centering
  \includegraphics[width=0.98\linewidth]{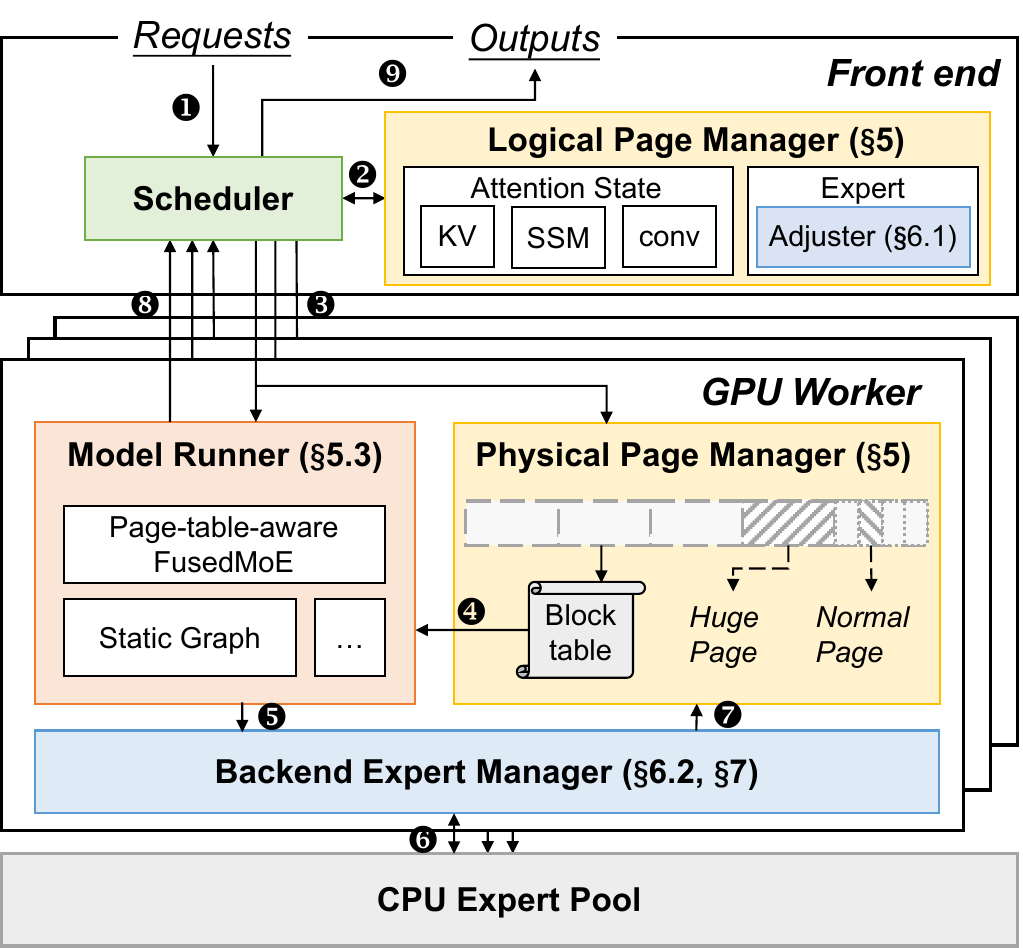}
  \caption{System overview of Rimuru.}
  \label{figure:overview}
\end{figure}

To address the above challenges, we present Rimuru, an MoE offloading serving system with unified memory management designed to maximize memory utilization. 
Rimuru combines unified memory management with workload-aware expert-cache adjustment, as shown in~\autoref{figure:overview}. The logical and physical page managers coordinate tensor-specific allocation over a unified shared page pool, while the model runner accesses tensors through page-table mappings. The adjuster determines changes in expert residency, and the backend expert manager integrates these changes with expert caching and asynchronous prefetching.

Specifically, upon request arrival, each request enters the queues managed by the scheduler (\raisebox{-0.15ex}{\ding{182}}). 
At each scheduling iteration, the logical page manager allocates attention-state pages, and the adjuster determines whether to expand or shrink the expert cache based on page availability and scheduling demand (\raisebox{-0.15ex}{\ding{183}}).
The scheduler dispatches the scheduled requests, page allocations, and cache-adjustment deltas to the workers (\raisebox{-0.15ex}{\ding{184}}).
The physical page manager maintains the corresponding mappings, which the model runner uses to access managed tensors during execution (\raisebox{-0.15ex}{\ding{185}}).
The backend expert manager combines predicted expert prefetches with cache-expansion deltas (\raisebox{-0.15ex}{\ding{186}}), transfers the required weights from CPU memory (\raisebox{-0.15ex}{\ding{187}}), and places newly cached experts in their assigned pages (\raisebox{-0.15ex}{\ding{188}}).
New cache mappings become visible only after loading completes; pages selected for eviction are reclaimed after the ongoing forward pass finishes. Execution results return to the scheduler (\raisebox{-0.15ex}{\ding{189}}), which streams generated tokens to the user (\raisebox{-0.15ex}{\ding{190}}).

\section{Unified Memory Management}

\subsection{Multi-view Memory Layout}

\begin{figure}
  \centering
  \includegraphics[width=\linewidth]{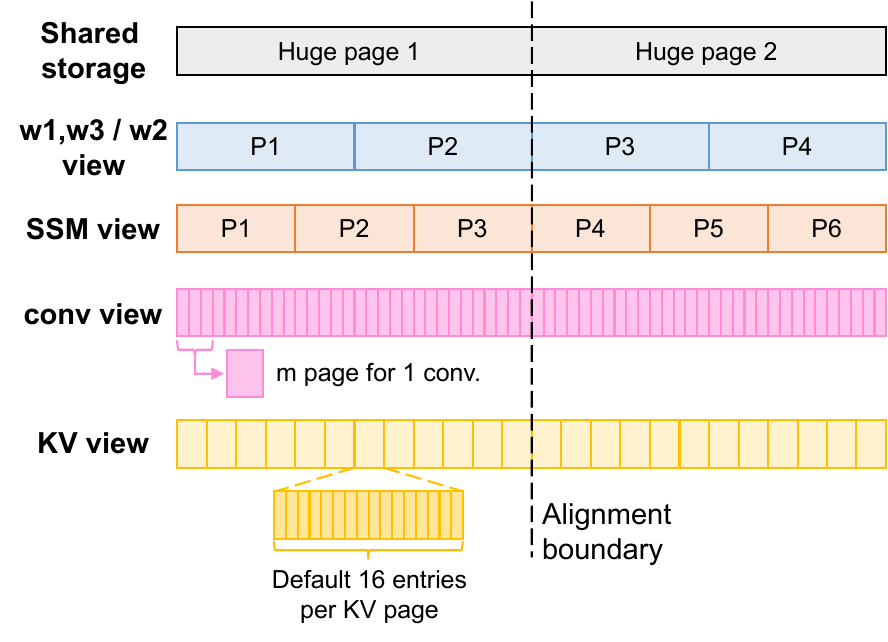}
  \caption{Multi-view memory layout of Rimuru. Each tensor type has its own tensor-specific view over the shared storage.}
  \label{figure:memory_layout}
\end{figure}

Rimuru allocates the available contiguous device memory as a shared page pool for heterogeneous tensors, and exposes tensor-specific memory views over this storage, as illustrated in~\autoref{figure:memory_layout}. 
Each view defines the layout and indexing of pages for a tensor type without reserving a separate memory region. Expert weights use different shape interpretations for the $w_1/w_3$ and $w_2$ matrices; attention states use KV, SSM, and convolutional-state views as required by the model. 
These views are different interpretations of the same shared pages. Once an allocation is safely reclaimed, its storage can be reassigned through another view.

To coordinate memory allocation across different tensor views, Rimuru aligns their page boundaries using a common huge page size. Since expert weights and SSM states typically require larger allocation units than other tensor types, Rimuru sets the huge page size to the least common multiple of their page sizes, denoted as $P_h = \mathrm{lcm}(P_e, P_s)$. For models without SSM states, Rimuru simply uses $P_h = P_e$. The page sizes of the KV cache and convolution states must also divide $P_h$.
Convolutional states require special handling because the size of a complete state may not divide the parent-page size. For example, in Qwen3.5-35B-A3B, each convolution state has a shape of $[3, 8192]$. Therefore, for a convolutional state of shape $[m, d]$, Rimuru uses a row of shape $[1, d]$ as the page allocation unit and allocates $m$ consecutive pages to store each convolution state, ensuring that individual pages can be aligned within huge pages.
For the KV cache, Rimuru follows PagedAttention, where each page stores $n$ KV entries, typically 16. The memory footprint of each page depends on the corresponding tensor shape, data type, and tensor-parallel configuration.

Based on this design, the front-end logical page manager consists of multiple specialized managers. Each manager is responsible for page allocation and reclamation for a specific memory view, and supplies page identifiers to the workers. These identifiers index the corresponding memory views, allowing computation to locate allocated tensor storage without repacking it into a separate layout. 
The mappings for expert and SSM views are straightforward. The KV cache mapping follows the existing design of paged attention. In contrast, the conv page table stores the start position of each allocation. A complete conv state is typically represented by a contiguous range of pages, such as indexed as $[\mathrm{start\_pos}, \mathrm{start\_pos} + m]$.

\subsection{Multi-level Page Allocation}

\begin{figure}
  \centering
  \includegraphics[width=\linewidth]{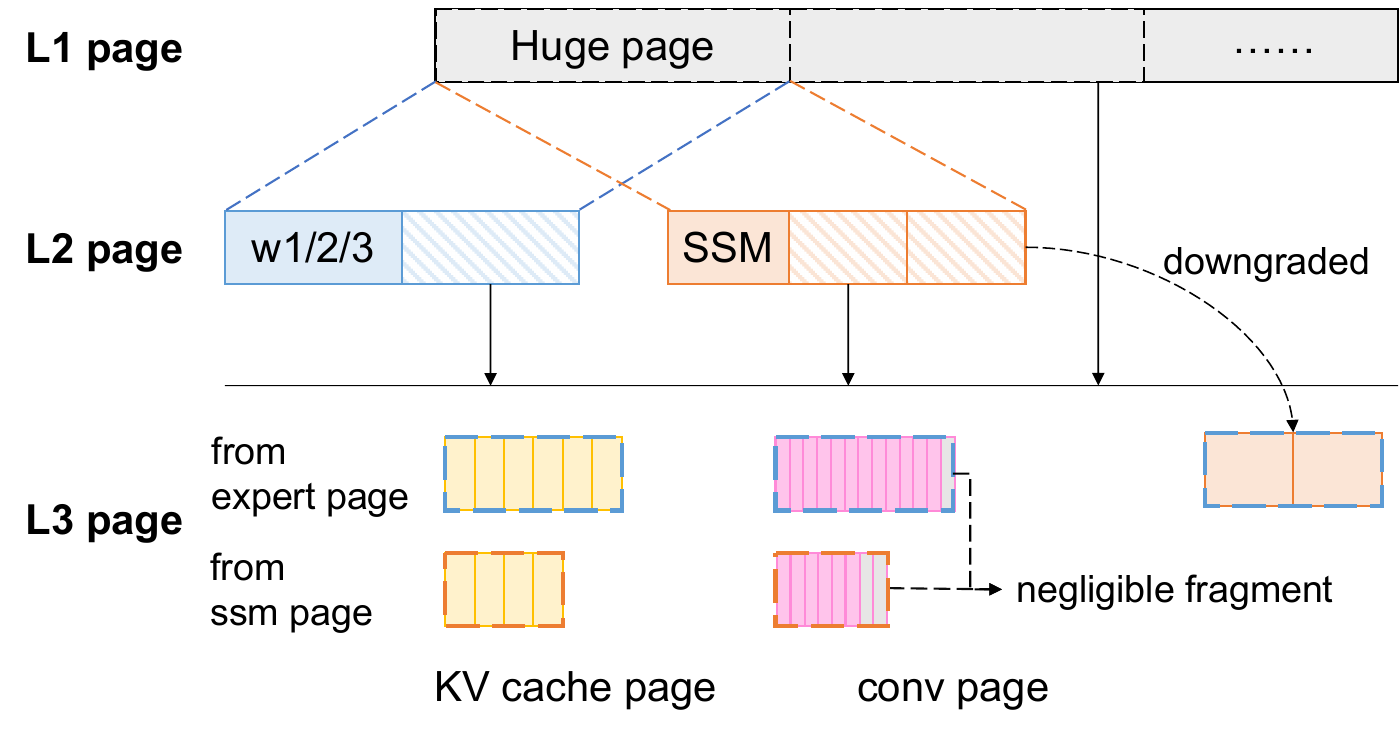}
  \caption{Multi-level page allocation and subdivision in Rimuru.}
  \label{figure:memory_allocate}
\end{figure}

Rimuru implements a multi-level page allocator that supports memory reuse across tensor types through on-demand splitting and reclamation while minimizing fragmentation, as illustrated in~\autoref{figure:memory_allocate}.
At the top level, the page pool maintains a queue of huge pages (L1 pages). Each huge page can be further partitioned into L2 pages, including expert pages and SSM pages, and L3 pages, including KV cache pages and conv pages. 
In addition, when the expert page size is an integer multiple of the SSM page size, $P_h=P_e$, and SSM pages can be allocated as sub-pages of expert pages, which simplifies management. 
The allocator supports two primary operations, allocation and reclamation.

Allocation proceeds through on-demand splitting. For expert and SSM requests, the system allocates the required number of pages from huge pages, and places the remaining pages into the free queue of L2 pages for future reuse. 
For KV and convolutional allocations, the system first reuses available sub-pages of the corresponding type and subdivides free L2 pages when necessary.
If insufficient L2 pages remains, it splits a free huge page into L2 pages, typically expert pages, and then allocates one of these L2 pages as L3 pages. 
This design avoids directly partitioning L1 pages into L3 pages, since L1 pages are much larger than L3 pages and such direct splitting may lead to fragmentation. 
As a result, the same physical storage can serve different tensor types over time, enabling elastic utilization across categories.
Across different requests, their L3 pages can be allocated from the same L2 page. However, Rimuru attempts to assign sub-pages from a single L2 page to the same request whenever possible to simplify reclamation.

Reclamation follows the reverse path of allocation. When an L3 page is released, the allocator updates the state of its parent page. Once all sub-pages of a parent page are freed, the allocator removes the entire sub-page range from the corresponding free set and restores the parent page as a free expert page or SSM page. A huge page returns to the global pool only when all intermediate pages derived from it have been fully reclaimed.


Through multi-level allocation and the merging of fully reclaimed sub-pages into larger allocation units, Rimuru prevents memory capacity from becoming stranded across tensor-specific regions. 
Internal fragmentation may still arise from partially utilized KV pages and unused tail space in convolution-state allocations. 
However, such fragmentation is bounded by the corresponding allocation granularities and remains small due to the fine-grained allocations for KV caches and convolution states. Therefore, memory fragmentation is negligible in Rimuru.

\subsection{Execution with Discrete Pages}

The computation of KV cache and Mamba states is already well described in prior systems~\cite{kwon2023pagedattention, zhang2025jenga}. Here, we focus on the execution of experts. 
Rimuru prepares expert weights through a dual-buffer design to overlap expert prefetching with execution , and extends the fused MoE kernel to access weights directly from cache pages or loading buffers.

\begin{figure}
  \centering
  \includegraphics[width=\linewidth]{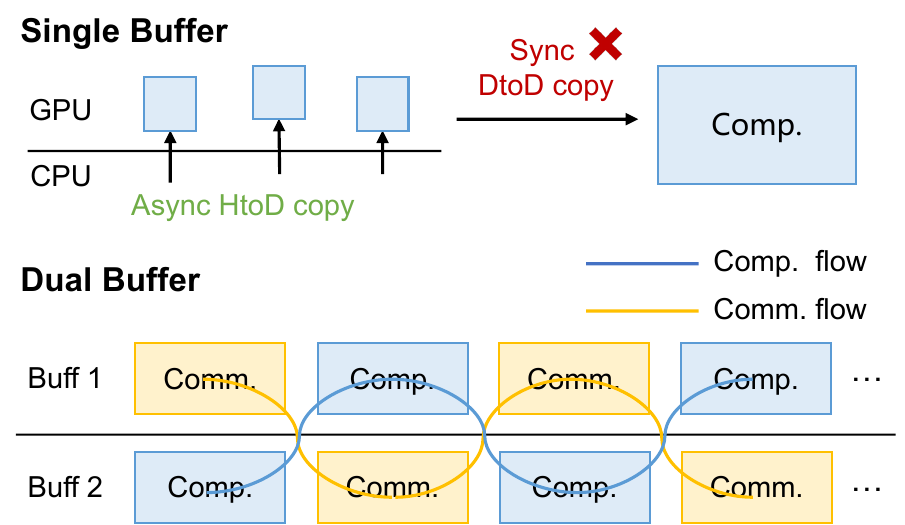}
  \caption{Single buffer vs. Dual buffers. Design for overlapping expert prefetch and execution, avoiding local DtoD copy.}
  \label{figure:dual_buffer}
\end{figure}

\textbf{Dual buffer.} 
Rimuru preallocates reusable loading buffers for non-resident experts, avoiding repeated page allocation and reclamation for temporary weight loading.
However, reusing the same buffer for current-layer execution and next-layer prefetching creates a read--write conflict: new transfers must wait for computation to finish, or weights must be staged elsewhere and copied through an additional local device-to-device transfer before execution.
To address this issue, Rimuru introduces a dual-buffer design, as illustrated in~\autoref{figure:dual_buffer}. 
It maintains two expert buffers, while one buffer supplies weights for the current layer, the other receives prefetched weights for the next layer. Their roles alternate across layers.
This design introduce a fixed memory overhead, which is twice the size of the expert layer. Rimuru therefore leverages the \texttt{expert\_offload\_limit} (see Section~6) to bound the buffer size. Since the number of experts that need to be prefetched or temporarily loaded at any time does not exceed this limit. 

\textbf{Page-table-aware fused MoE kernel.}
Conventional fused MoE operators index a contiguous expert-weight array. In Rimuru, cached weights occupy separate pages, while non-resident experts are loaded into temporary buffers. Gathering these weights into a packed array would add device-to-device copies. Rimuru instead uses expert-location metadata to compute directly from cache pages or loading buffers.

The operator takes input activations, routing results, and metadata identifying each expert's weight source and location. Cached projections are located through page identifiers, while buffered projections are located through buffer slots. For the joint $W_1/W_3$ projection, output columns select their corresponding projection pages without physically concatenating the weights. Let $d$ and $f$ denote the local hidden and intermediate dimensions. Each cached projection occupies a dense $[f,d]$ page. For expert $e$, output column $n \in [0,2f)$ and reduction offset $t \in [0,d)$ address the weight element as
\begin{equation}
\begin{aligned}
p_e(n) &=
\begin{cases}
p_{e,1}, & n<f,\\
p_{e,3}, & n\geq f,
\end{cases}\\
\operatorname{addr}_e(n,t)
&= B_{\mathrm P}+p_e(n)fd+(n\bmod f)d+t,
\end{aligned}
\end{equation}
where $B_{\mathrm P}$ is the page-pool base and $p_{e,1}$ and $p_{e,3}$ are the projection-page identifiers; offsets count stored weight elements. Routed token--expert assignments are grouped by expert, and each GPU program computes an output tile while reusing its weight tile across input rows. Page lookup and source selection occur before the GEMM reduction loop, which loads dense tiles and advances pointers using fixed strides. The additional indirection is therefore resolved per output tile rather than at every reduction step. After activation and the $W_2$ projection, expert outputs are combined using routing weights.

For execution chunks with at most 32 assignments and no expert remapping, the GEMMs form expert groups internally, avoiding separate alignment launches and intermediate alignment metadata. Assignments to each expert are ordered by occurrence; every $B_M$-th occurrence, starting at rank zero, owns up to $B_M$ matching rows, while other candidates return before loading weights. Larger workloads use explicit alignment to avoid repeated local grouping. The final reduction writes directly into the output tensor, avoiding a preliminary zero fill and a workspace-to-output copy.

\textbf{Support for piecewise CUDA graph.} 
This layout is compatible with piecewise CUDA Graph execution for expert computation, since steady-state data structures have fixed shapes and stable memory addresses. The original page storage, expert page table, dual buffers, and auxiliary metadata tensors are allocated once and reused across steps. At runtime, only tensor contents are updated. As a result, the captured graph reflects a stable tensor topology, while page residency and request-dependent routing remain dynamic via indirection tables.


\section{Workload-aware Expert Cache Adjuster}

To flexibly utilize unified memory, Rimuru further introduces a workload-aware expert cache adjuster. 
The adjuster balances attention-state capacity against expert-loading overhead. It shrinks the expert cache when attention-state allocation requires additional pages and expands it when sufficient capacity becomes available, while limiting excessive eviction and repeated loading.
The adjuster separates control from execution. The front-end logical expert page manager determines whether the cache should expand or shrink based on the current page availability and sends the resulting cache delta to the worker. On the worker side, the expert manager incorporates this delta into the existing prefetch path and updates page table.

\textbf{Hotset-guided Cache Layout.} Given that MoE models exhibit skewed expert access patterns~\cite{li2023accelerating, fang2025klotski} and that online statistics introduce extra overhead, the adjuster relies on an offline expert hotset profile. For each MoE layer, this profile stores a complete ordering of experts together with their empirical routing probabilities. Rimuru uses this ordering as a prefix-based cache policy. 
During both initialization and cache resizing, experts are selected according to this order. The empirical routing probabilities are further used to compare priorities across layers when adjusting cache sizes.

At initialization, the first MoE layer has no prefetch window and therefore caches all experts. Subsequent MoE layers, referred to as adjustable layers, cache only a prefix of their hotset ordering. By default, Rimuru initializes a large expert cache within the available page budget. 
During adjustment, each layer enforces a unified $expert\_offload\_limit$ as a bound on the number of offloaded experts, which prevents excessive offloading and avoids latency degradation.

\subsection{CPU-side Adjustment Policy}

On the CPU side, the control logic consists of two main decisions, whether to resize the cache and which experts to load or evict. This decision is made after the scheduler completes request scheduling, since the remaining available pages reflect the current short-term memory pressure. Then, the adjuster constructs an \texttt{ExpertCacheDelta}, which includes cache delta set, and the reserved page identifiers for newly loaded experts, and sends it to the worker by scheduler. The high-level workflow of the CPU-side adjustment policy is shown in \autoref{alg:expert-cache-adjust}.

\begin{algorithm}[t]
  \caption{Control-side expert cache adjustment}
  \label{alg:expert-cache-adjust}
  \small
  \begin{algorithmic}[1]
  \Require Adjustable layers $\mathcal{L}$, hotset rankings $\{H_l\}$, routing probabilities $\{p_l\}$, cache sizes $\{C_l\}$, expert\_offload\_limit $C_{\min}$, free expert blocks $F_b$, required shrink blocks $R_b$
  \Ensure ExpertCacheDelta $\Delta$ or $\emptyset$
  \If{$\mathcal{L} = \emptyset$}
      \State \Return $\emptyset$
  \EndIf
  \If{$R_b > 0$}
      \State $N \gets \left\lceil (R_b + 1) / 3 \right\rceil$
      \State $\Delta S \gets \Call{GreedyShrink}{N, \{C_l\}, C_{\min}, \{p_l\}}$
      \State \Return \Call{ExpertCacheDelta}{$[\;], \Delta S$}
  \ElsIf{$F_b < 2$}
      \State $\Delta S \gets \Call{GreedyShrink}{1, \{C_l\}, C_{\min}, \{p_l\}}$
      \State \Return \Call{ExpertCacheDelta}{$[\;], \Delta S$}
  \Else
      \State $N \gets \min\!\left(5,\left\lfloor \max(0, F_b - 9) / 3 \right\rfloor\right)$
      \State $\Delta E \gets \Call{GreedyExpand}{N, \{C_l\}, \{p_l\}}$
      \State \Return \Call{ExpertCacheDelta}{$\Delta E, [\;]$}
  \EndIf
  \end{algorithmic}
\end{algorithm}

Specifically, the trigger for cache resizing is based on the remaining capacity of expert pages, including those that can be obtained by splitting huge pages. When waiting requests exist, the system prioritizes shrinking according to demand. Let $F_b$ denote the number of currently free expert pages, and let $R_b$ denote the number of additional expert pages required by the scheduler to satisfy the current scheduling result. If $R_b > 0$, the adjuster converts this shortage into the number of experts to shrink: $N_{\mathrm{shrink}} = \left\lceil \frac{R_b + 1}{3} \right\rceil$, where the divisor of 3 accounts for the three expert pages occupied by each expert, one for each projection matrix.
It then attempts to select up to $N_{\mathrm{shrink}}$ experts, subject to the per-layer cache lower bound, and generates the corresponding shrink delta (Line 4-7). The additional page in the numerator provides a small buffer for the next adjustment. If no explicit shortage exists but $F_b < 2$, the adjuster still attempts to shrink one expert (Line 8-10). Otherwise, the adjuster considers expansion.

During expansion, the system reserves capacity for three experts. 
Since each expert occupies three pages for its weights, expansion is triggered when $F_b > 9$. This gap between the shrink threshold (2) and the expand threshold (9) is intended to reduce frequent reversals as memory availability fluctuates.
The expansion budget is computed as $N_{\mathrm{expand}} = \min\left(5,\left\lfloor \frac{\max(0, F_b - 9)}{3} \right\rfloor\right)$.
The cap of five experts bounds the expansion per forward pass, limiting the additional loading work introduced into the prefetch pipeline (Line 11-14).

Target selection for cache resizing follows a greedy policy guided by the hotset prefix (Line 6, 9, 13). For \textsc{GreedyExpand}, each adjustable layer proposes the next uncached expert in its hotset ordering $\{H_l\}$. The adjuster selects the $N_{\mathrm{expand}}$ experts with the highest routing probabilities $p_l(e)$. \textsc{GreedyShrink} decisions follow a symmetric rule. Each layer proposes the last expert in its current prefix. 
Layers whose cached prefix length has reached $C_{\min}$ are excluded from shrinking.
The cost of removing an expert is defined by its routing probability $p_l(e)$, and the adjuster selects one or $N_{\mathrm{shrink}}$ experts with the smallest cost.


\subsection{GPU-side Delta Application}

On the GPU side, the expert manager applies the cache delta within the forward pipeline. Expanded experts are merged into the prefetch queue of the next layer. Before layer $l$ begins execution of experts, the expert manager schedules asynchronous prefetching for layer $l+1$. The prefetch queue contains both predicted experts and the expand delta associated with layer $l+1$. The manager explicitly removes duplicates between these two sources to avoid redundant copies. 
Experts from the expand delta are issued with higher priority and must be completed before execution.
When execution reaches layer $l+1$, the manager checks whether there are pending expand deltas for that layer. If so, it waits for the corresponding copy events and then updates the page table. This ordering, where copies are completed before page table updates, ensures correctness.


Shrink is simpler and only requires invalidating the corresponding page table entries while controlling the timing. When the GPU receives a cache delta with experts to evict, it does not immediately free their pages, but waits until the current forward pass completes. The backend manager keeps these experts visible during the ongoing forward pass to increase reuse. After the forward pass, the manager commits updates to the page table. The frontend then reclaims the old pages and returns them to the page pool.




\section{Implementation}

Rimuru is implemented on top of vLLM with approximately 4,400 lines of Python code and 800 lines of C++ code. The control logic and memory management are implemented in Python, the page-table-aware expert GEMM kernels in Triton, and the expert-loading functionality through C++ extensions. By building on vLLM, Rimuru supports a broad range of modern LLMs compatible with vLLM.

\textbf{Memory views and integration.}
Rimuru allocates the shared page pool as a contiguous device byte buffer. It uses \texttt{.view(dtype)} to reinterpret the element type and \texttt{.view(shape)} to construct tensor-specific layouts over the same storage without copying data. The block manager returns typed page identifiers that index these views on each worker. For attention states, Rimuru adapts vLLM's block tables and state indices to the shared layout while retaining the underlying attention and state-update computations. Convolutional-state mappings additionally record the starting row-page index of each consecutive allocation. These adaptations connect the shared allocator to the existing computation interfaces.

\textbf{Expert management and prefetching.}
The backend expert manager maintains host-resident weights, device cache mappings, loading buffers, and prefetch state. Host weights are stored in pinned memory for asynchronous transfers. Rimuru adopts the expert-prediction strategy from Fate~\cite{fang2026fate} and passes predicted expert identifiers to a background prefetch worker. The worker filters cached experts and coordinates predicted loads with cache-expansion targets. Actual routing results determine the weights required for execution, with unavailable experts loaded before computation. The model runner supplies their locations to the page-table-aware fused MoE operator. Moreover, most expert prediction strategies developed in other expert offloading systems remain orthogonal to Rimuru.

\textbf{Extensibility.}
The shared storage abstraction accommodates weight formats such as BF16 and INT8 through dtype-specific views. Adapting the memory manager to another format primarily requires updating the view's element interpretation and mapping calculations. Models composed of supported tensor types reuse the existing views. New  model or heterogeneous tensor types can be incorporated into the unified memory manager by adding views that define their storage layouts, allocation units, and index mappings.

\textbf{Tensor Parallelism.}
For multi-device execution, Rimuru adopts tensor parallelism to balance memory occupancy and expert-weight transfers. Under identical per-rank configurations, workers retain the same expert-residency layout and load equal-sized shards of the same selected experts. Each adjustment therefore releases or consumes the same capacity on every device and introduces the same expert I/O volume. This symmetry allows the shared allocation and cache-adjustment policy to operate consistently across workers without introducing uneven expert-loading pressure.

\section{Evaluation}

\subsection{Evaluation Setup}

\textbf{Hardware.} 
We evaluate Rimuru and baselines on two hardware configurations, and conduct a comprehensive evaluation across single-, dual-, and eight-GPU settings to demonstrate that the effectiveness of Rimuru scales with system size. 
One server is equipped with eight NVIDIA RTX 5090 GPUs (32~GB each), two Intel Xeon Platinum 8470Q CPU, and 720~GB of host memory, with PCIe~5.0 $\times$16 interconnects between CPU and GPU. The other server is equipped with two NVIDIA RTX 4090D GPUs (24~GB each), one Intel Xeon Platinum 8368Q CPU, and 256~GB of host memory, with PCIe~4.0 $\times$16 interconnects between CPU and GPU.

\textbf{Models.} 
We use MoE models from the Qwen family, which provide a range of models at different scales with strong performance. The Qwen-3 series consists of full-attention models, while the Qwen-3.5 series adopts hybrid attention. Specifically, we evaluate Qwen3-30B-A3B, Qwen3.5-35B-A3B, and Qwen3.5-122B-A10B.

\textbf{Workload.}
To simulate real-world serving workloads, we sample inputs from ShareGPT~\cite{ShareGPT}, a widely used dataset of real user interactions with ChatGPT. 
As such, it serves as a representative benchmark for LLM serving, capturing practical usage scenarios, workload diversity, and variability in request patterns.

\begin{figure*}
  \centering
  \includegraphics[width=\linewidth]{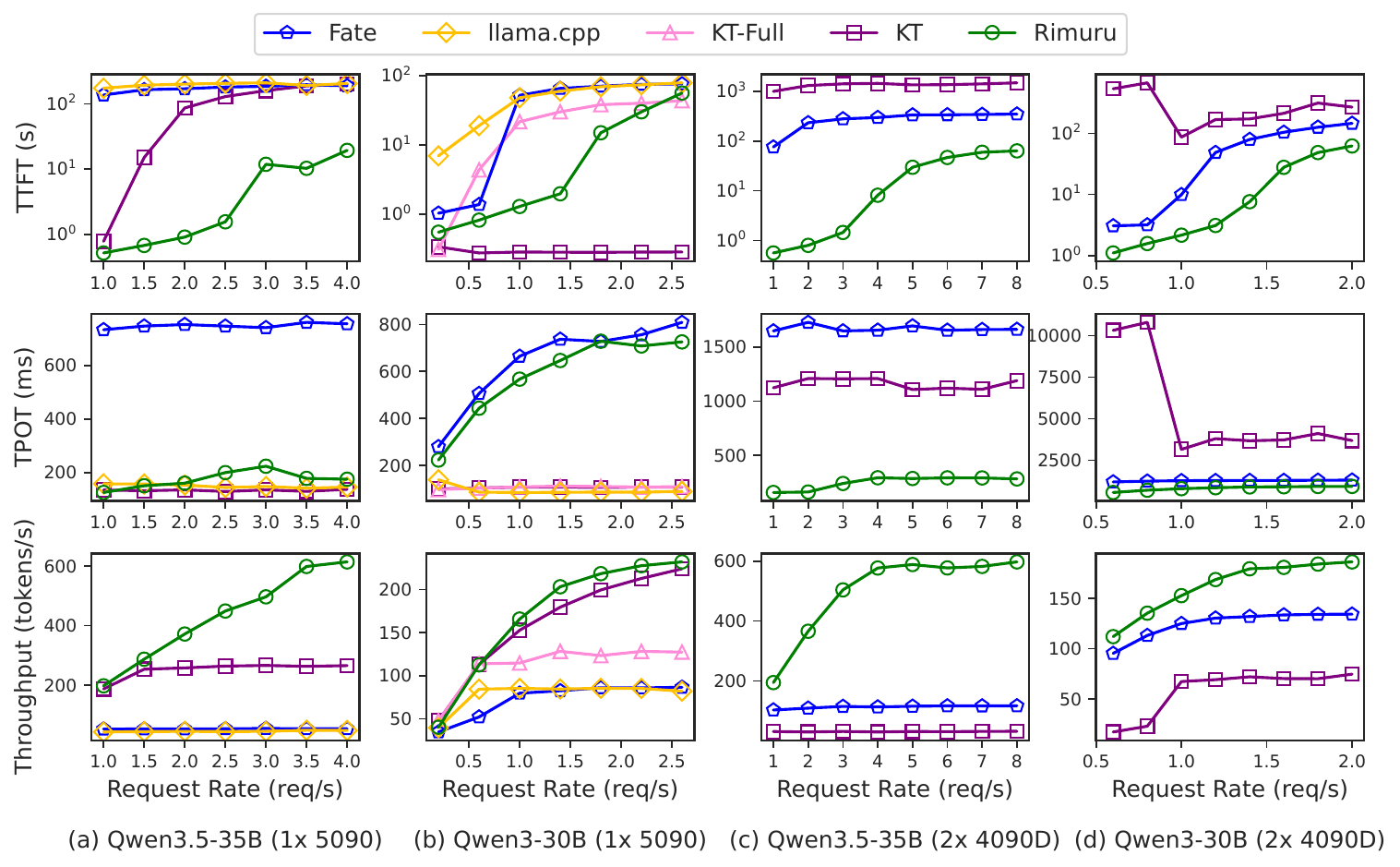}
  \caption{Comprehensive serving evaluation of expert offloading. Columns show the four model and hardware configurations, and rows report TTFT, TPOT, and output-token throughput as request rate increases.}
  \label{figure:tp_offloading}
\end{figure*}

\textbf{Baselines.}
We evaluate Rimuru under two resource-limited settings: one where GPU memory accommodates all model weights with little headroom, and one where memory is insufficient, requiring partial expert offloading.
In the former setting, we compare Rimuru with the following popular serving systems to show that Rimuru improves throughput by enabling transient concurrency:
\begin{itemize}[leftmargin=*, itemsep=0pt, topsep=0pt, parsep=0pt]
\item {\textbf{vLLM}}~\cite{kwon2023pagedattention}: a high-throughput LLM serving engine that mitigates memory fragmentation via paged attention.
\item {\textbf{SGLang}}~\cite{zheng2024sglang}: a structured generation framework that leverages RadixAttention to reuse KV cache.
\end{itemize}
In the latter setting, we compare Rimuru with following offloading systems to highlight the benefit of dynamically adjusting the expert cache over static configurations:
\begin{itemize}[leftmargin=*, itemsep=0pt, topsep=0pt, parsep=0pt]
\item {\textbf{KTransformers}}~\cite{chen2025ktransformers}: a flexible framework for heterogeneous LLM inference that leverages Intel AMX instructions to accelerate CPU-side matrix computation.
\item {\textbf{Fate}}~\cite{fang2026fate}: an MoE offloading system that reduces I/O overhead through high-hit-rate expert prefetching and caching. We modify Fate to adapt it to Qwen models.
\item {\textbf{llama.cpp}}~\cite{llama}: a lightweight C/C++ implementation for running LLMs on CPUs or resource-constrained devices.
\end{itemize}
Note that most existing offloading systems do not support CUDA Graph. Since the performance gains from CUDA Graph may bias the comparison of system designs, we disable CUDA Graph in all systems when supported.

\textbf{Metrics.}
The primary metric targeted by Rimuru is output token throughput (tokens/s), which reflect overall generation speed. In addition, as a serving system, we also consider TTFT and TPOT.

\subsection{End-to-end Performance}

\textbf{Compare with offloading systems.} 
\autoref{figure:tp_offloading} shows the evaluation of KTransformers, Fate, llama.cpp, and Rimuru. We serve Qwen3-30B and Qwen3.5-35B on a single 5090 GPU and two 4090D GPUs. Rimuru achieves up to 1.39$\times$--19.03$\times$ throughput and reduces TTFT by up to 99\% compared with most baselines, except for a smaller gap with KT. Although Rimuru exhibits higher TTFT in some cases, it supports higher concurrency through the expert cache adjuster, which leads to superior overall throughput.

Fate represents typical MoE offloading systems on edge devices, which focus on expert prediction and caching to hide I/O overhead. However, they overlook conflicts when integrating with modern serving engines. Strict memory isolation leads to underutilization of GPU memory. Rimuru addresses this issue by breaking memory isolation and adopting a flexible cache adjustment strategy, which enables switching between high throughput and low latency. 
Finally, Rimuru achieves up to 11.34$\times$ the throughput of Fate.

In contrast, KTransformers and llama.cpp avoid expert movement and instead execute MoE computation on CPU using optimized kernels. KTransformers outperforms llama.cpp due to its MoE kernel based on Intel AMX instructions. 
Rimuru also consistently outperforms llama.cpp, achieving up to 12.7$\times$ its throughput and reducing TTFT by 90.50\%--99.7\%. 
Since llama.cpp only supports single-GPU execution, it is not included in (c) and (d).

On the 4090D platform, the Intel Xeon Platinum 8368Q CPU does not support AMX instructions, so we configure KTransformers to use AVX2 for CPU-side expert computation. As a result, its performance is significantly degraded. 
Under high workload, Rimuru achieves up to 19.03$\times$ the throughput of KTransformers on the 4090D platform.
On the 5090 platform, which supports AMX instructions, KTransformers shows strong performance on Qwen3. Since KTransformers uses a fixed, manually configured expert cache size, we evaluate two settings: KT-Full, where the expert cache fully occupies GPU memory, and KT, where some memory is reserved for KV cache and Mamba states.
KT-Full exhibits worse performance due to excessive expert caching that reduces KV cache capacity, limits concurrency, and causes request accumulation. 
In contrast, Rimuru dynamically adjusts the expert cache and reallocates memory to the KV cache, which increases concurrency and improves throughput.
On the other hand, KT achieves lower TTFT and TPOT than Rimuru on Qwen3 by leveraging high-performance CPU kernels without transferring expert weights. However, Rimuru achieves higher throughput by supporting greater request concurrency through elastic expert caching.
Results across two platforms also show that KTransformers relies heavily on advanced CPU instruction sets, which are often unavailable on older servers and PCs in resource-constrained environments.
For Qwen3.5, the expert size is one-third smaller than that of Qwen3, reducing the I/O pressure. As a result, Rimuru significantly outperforms KTransformers.
Overall, Rimuru achieves up to 2.30$\times$ the throughput of KTransformers on the 5090 platform.

\begin{figure}
  \centering
  \includegraphics[width=0.98\linewidth]{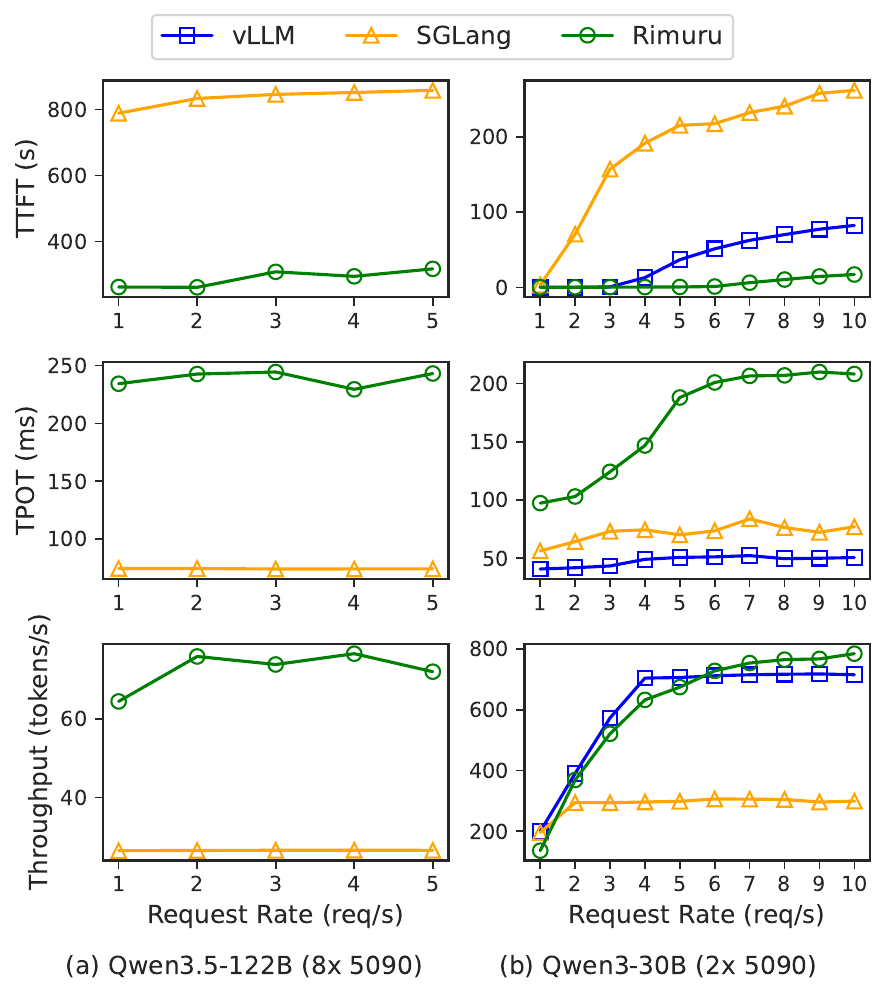}
  \caption{Serving evaluation of SGLang, vLLM, and Rimuru.}
  \label{figure:tp_allin}
\end{figure}

\textbf{Compare with serving systems.} \autoref{figure:tp_allin} shows the evaluation of SGLang, vLLM, and Rimuru when serving Qwen3.5-122B on eight 5090 GPUs and Qwen3-30B on two 5090 GPUs. Rimuru achieves 1.1$\times$--2.88$\times$ throughput compared with SGLang and vLLM. And Rimuru reduces TTFT by 63\%--99\%. These results demonstrate the effectiveness of Rimuru and highlight the necessity of expert offloading when deploying MoE models under resource constraints.

The improvement stems from better handling of high workload under limited memory. 
By admitting more concurrent requests under memory pressure, Rimuru alleviates request accumulation and reduces queueing delays, which contributes substantially to the observed TTFT improvements.
Under low workload, Rimuru shows limited degradation. It dynamically increases the number of cached experts, which reduces I/O overhead. For example, Rimuru achieves roughly half the TPOT under low workload compared to high workload. Although Rimuru has higher TPOT than the baselines, the overall request completion time remains lower.


We omit vLLM from panel (a) because the evaluated configuration encounters an OOM error during Triton GDN kernel autotuning. This limitation highlights the necessity of offloading systems, which provide flexibility across limited deployment environments. Since Rimuru is built on top of vLLM, it is also affected. Nevertheless, Rimuru still demonstrates strong performance.





\subsection{Ablation Study}

\begin{figure}
  \centering
  \includegraphics[width=0.98\linewidth]{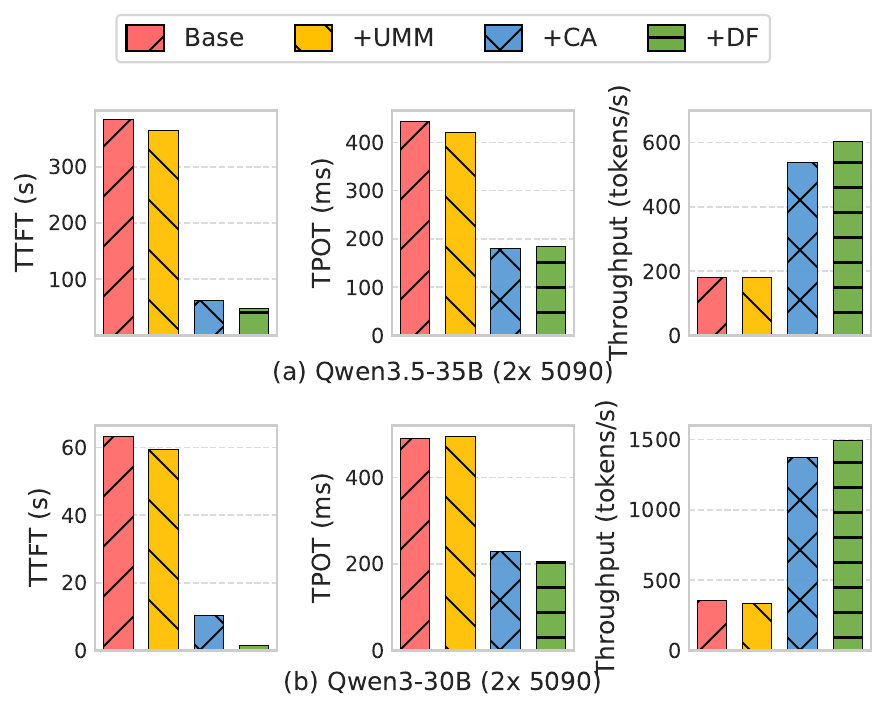}
  \caption{Ablation study of Rimuru. UMM: Unified Memory Management, CA: Cache Adjuster, DF: Discrete FusedMoE.}
  \label{figure:ablation}
\end{figure}

To validate the effectiveness of each component in Rimuru, we conduct an ablation study. Starting from Fate as baseline, we progressively add unified memory management, expert cache adjustment, and the page-table-aware fused MoE kernel. The experiments are conducted on Qwen3.5-35B and Qwen3-30B using two 5090 GPUs, which correspond to the cases where offloading is required and not required, respectively. The results are shown in~\autoref{figure:ablation}.

Specifically, unified memory management provides the underlying mechanism of memory sharing rather than directly improving serving performance. 
Consequently, its performance remains comparable to the baseline. 
The benefit of unified allocation emerges when combined with the workload-aware cache adjuster.
By elastically adjusting expert residency in response to memory pressure to improve memory utilization, the adjuster increases throughput to approximately 3.85$\times$ that of the baseline and reduces TTFT by 83.9\%.
Finally, the page-table-aware FusedMoE kernel eliminates copying cached experts into contiguous execution buffers, further increasing throughput to 4.19$\times$ and reducing TTFT by approximately 97.8\%.
During prefill, parallel processing of many tokens activates most experts, making weight copying particularly expensive. Our kernel avoids these copies by directly accessing expert weights from the shared page pool.
During decoding, fewer experts are activated, reducing device-to-device copy overhead and thus yielding smaller TPOT improvements.

\subsection{FusedMoE Kernel Efficiency}

\begin{figure}[t]
  \centering
  \includegraphics[width=0.97\columnwidth]{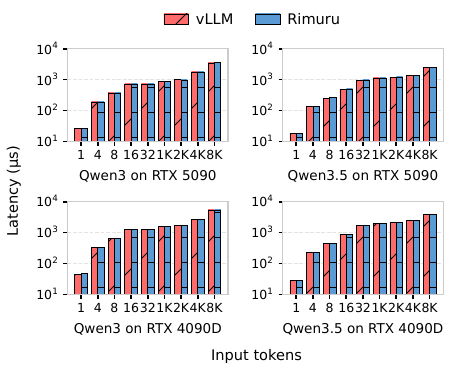}
  \caption{Comparison of FusedMoE kernels. }
  \label{fig:discrete-kernel-latency}
\end{figure}

We compare the page-table-aware FusedMoE kernel with the original contiguous-weight kernel of vLLM.
We evaluate the two models' expert configurations on both GPU platforms, using small token batches ($M=1$--32) and large token batches ($M=1024$--8192) to represent decoding- and prefill-sized inputs, respectively.
Both kernels use identical inputs, routing decisions, and weights.

Across all results in Figure~\ref{fig:discrete-kernel-latency}, our kernel increases operator latency by 0.83\% on average, indicating that our design keeps the net cost of non-contiguous execution small. For large token batches ($M\ge1024$), the mean paired overhead remains within 1.31\% in every tested configuration, consistent with address-mapping costs being better amortized by batched expert computation. Small token batches show greater relative variation, with a maximum increase of 9.63\% in the single-token case. However, the absolute latency increase remains small, reaching at most 3.38~$\mu$s. These results show that our kernel preserves near-baseline resident-compute performance while directly consuming weights from the shared page pool.
Moreover, all activated experts reside in the page pool. We also evaluate mixed expert placements across the page pool and loading buffer. Varying the fraction of page-resident experts from 50\% to 100\% changes the mean overhead by at most 0.22\%. We omit these results for brevity.

\subsection{Sensitivity Analysis}

\begin{figure}
  \centering
  \includegraphics[width=\linewidth]{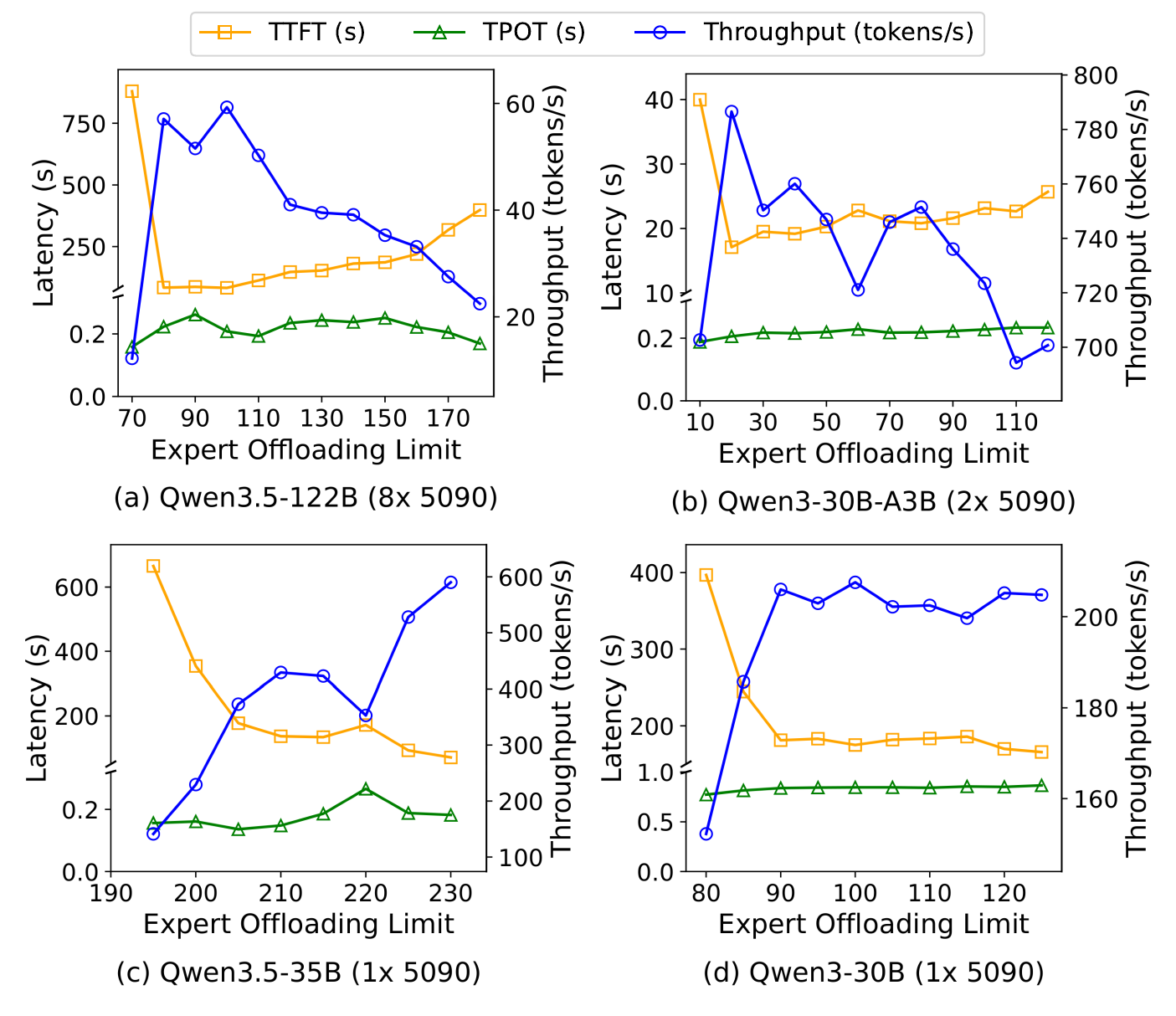}
  \caption{Impact of \texttt{expert\_offload\_limit}.}
  \label{figure:best_limit}
\end{figure}

In Rimuru, the optimal value of \texttt{expert\_offload\_limit} (denoted as EOL) is difficult to determine, as it varies with model architecture, GPU memory capacity, and workload. We therefore analyze the impact of \texttt{EOL} on different metrics, as shown in~\autoref{figure:best_limit}. The starting point of the x-axis in each subfigure indicates the minimum number of experts offloaded. 

Overall, across GPU memory capacities and model sizes, throughput, TTFT, and TPOT show consistent trends with respect to \texttt{EOL}. At the minimum \texttt{EOL}, Rimuru degenerates to a fixed-expert offloading system similar to Fate. Cached experts occupy most GPU memory, leaving little space for KV cache and Mamba states. Moreover, the \texttt{EOL} constraint prevents further offloading to release memory, leading to request accumulation, very high TTFT, and low throughput.
However, slightly increasing \texttt{EOL}, for example by about ten experts, significantly improves throughput and reduces TTFT. Peak throughput and the best \texttt{EOL} are typically achieved in this range. This is because expert weights are large, and offloading about ten experts frees enough memory to substantially increase concurrency. Further increasing \texttt{EOL} can still yield good performance, but excessively large values are suboptimal, as shown in (a) and (b). Under bursty workloads, many experts may be offloaded, and when load decreases, the system requires a long recovery period to reload them, degrading throughput.
In contrast, TPOT remains largely unaffected. Due to the fine-grained expert structure, only a small fraction of parameters is activated during decoding, and most accesses are covered by prefetching and the hotset-based cache policy. As a result, TPOT stays stable across \texttt{EOL} settings.

\section{Related Work}

\textbf{Memory management for LLM serving.}
High-throughput serving systems have invested heavily in managing the growing KV cache, because it often dominates GPU memory at scale~\cite{qin2026mooncake, gao2025apt, agrawal2024taming}.
Early LLM serving systems such as Orca~\cite{yu2022orca} highlighted the importance of iteration-level scheduling and continuous batching for autoregressive decoding.
vLLM~\cite{kwon2023pagedattention} then introduce PagedAttention, which can substantially improve memory efficiency and throughput. This design underpins many production engines.
SGLang~\cite{zheng2024sglang} expands the design space with structured multi-call programs and reuse KV via RadixAttention.
More recently, Jenga~\cite{zhang2025jenga} addresses heterogeneity across layers and embedding layouts with attention-aware allocation layered on top of paged KV management.
These lines of work substantially improve KV efficiency, but for model weights, they typically follow parallelism strategies to place them in device memory without additional management.
In contrast, Rimuru eliminates memory isolation in MoE serving by jointly managing cached expert weights and attention states within a unified memory framework, enabling dynamic memory sharing between expert residency and attention-state allocations.

\textbf{MoE inference and expert offloading.}
Most existing MoE offloading systems focus on expert prediction and caching~\cite{hwang2024pregatedmoe, tang2026moe}. MoE-Infinity~\cite{xue2024moe} proposes activation-aware prefetching and caching, while Fate~\cite{fang2026fate} improves expert hit rate through cross-layer prediction and a shallow-favoring expert cache. Similar designs are adopted in Mixtral-offloading~\cite{eliseev2023fast} and fMoE~\cite{yu2025fmoe}. These systems are largely complementary to ours. 
Furthermore, Fiddler~\cite{kamahori2025fiddler} and KTransformers~\cite{chen2025ktransformers} explore CPU--GPU co-execution. 
Another line of work targets offline workloads. MoE-Lightning~\cite{cao2025moe} and Klotski~\cite{fang2025klotski}improve throughput by reshaping execution to better overlap communication and computation.
In contrast, Rimuru focuses on how much capacity expert caching should retain as request-state demand changes and how the released capacity becomes reusable. Rimuru unifies experts and attention states and dynamically adjusts the expert cache online. Expert prediction and prefetching provide complementary mechanisms for reducing the loading overhead.

\section{Conclusion}
In this paper, we presented Rimuru, a modern LLM serving system that employs unified memory management to improve throughput in resource-constrained environments. Rimuru introduces tensor-specific memory views and a multi-level allocator to support memory allocation and reuse across different granularities. A page-table-aware fused MoE operator directly consumes cached and buffered weights, avoiding copies into a packed expert array. A workload-aware adjuster changes expert residency in response to attention-state pressure and coordinates these changes with asynchronous loading and execution. Together, these mechanisms allow memory capacity to follow changing serving demands while accounting for expert-loading overhead. 
Implemented on vLLM, Rimuru achieves up to 19.03$\times$ the throughput of the baselines.

\bibliographystyle{ACM-Reference-Format}
\bibliography{sample-base}

\end{document}